\documentstyle[multicol,aps,prl,epsf,psfig]{revtex}
\begin{document}
\title{Tapping Spin Glasses}
\author{David S. Dean and Alexandre Lef\`evre\\
{IRSAMC, Laboratoire de Physique Quantique, Universit\'e Paul Sabatier, 118 route de Narbonne, 31062 Toulouse Cedex}
}
\date{January 10 2001}
\maketitle

\begin{abstract}
We consider a tapping
dynamics, analogous to that in experiments on granular media, 
on spin glasses and ferromagnets on random thin graphs.
Between taps, zero temperature single spin flip dynamics takes the
system to a metastable state.
Tapping, corresponds to flipping simultaneously any spin with 
probability $p$. This dynamics leads to a
stationary regime with a steady state energy $E(p)$. We analytically
solve this dynamics for the  one dimensional ferromagnet
and $\pm J$ spin glass. Numerical simulations for 
spin glasses and ferromagnets of higher connectivity are carried out, 
in particular we find a novel first order transition for the ferromagnetic 
systems.
\vskip 0.5cm


\end{abstract}

\begin{multicols}{2}
Complex systems such as granular media and spin glasses
have an exponentially large number $N_{MS}$ of
metastable states, also called blocked or jammed 
configurations, {\em i.e.} $N_{MS} = \exp(Ns_{Edw})$, where  $s_{Edw}$ is the
Edwards entropy per particle \cite{edw}.  
In granular media the thermal energy
available is not sufficient to allow the rearrangement of a single 
particle and hence the system is effectively at zero temperature in
the thermal sense and when not  perturbed are stuck in a metastable state.
Edwards has proposed \cite{edw} that if one slightly perturbs
such systems by external forcing such as tapping, the asymptotic measure
over the metastable states satisfying the appropriate macroscopic
constraints is flat, this idea has recently been further investigated
\cite{jorge,nico}. 
In a, now classic, experiment on granular media
\cite{exps} dry hard soda glass spheres  placed in a glass tube are
tapped by using a piston to move
the tube vertically 
through a sine cycle. The tapping parameter $\Gamma$ is defined to 
be the ratio of the maximal acceleration due to the piston 
in the cycle to $g$ the acceleration
due to gravity. After an initial irreversible curve obtained by increasing
the tapping amplitude slowly, the system arrives on a reversible curve where
the density is a monotonic function of $\Gamma$, the highest packing
densities being obtained at lowest tapping rate. 
Numerical simulations on granular media \cite{mehta1} reveal similar behavior.
There has been a considerable theoretical effort to understand 
the dynamics of granular media, especially the slow relaxation towards 
the final asymptotic density under tapping, both theoretically and 
by simulations on various models \cite{mehta1,mehta2,Ben,GM}. Our
aim is to gain some understanding of the stationary regime of tapped
systems such as granular media.
 
The 
possibility of using spin glasses as a paradigm for granular material
was first suggested in \cite{edme}, and recently it has been shown in
\cite{dean2,lefde} that spin glasses and ferromagnets on random thin graphs
have an extensive Edwards entropy with respect to single spin flip dynamics. 
In this paper we study a very natural tapping dynamics on these systems:
We allow the system to
evolve under a random  zero temperature single spin flip 
dynamics where
only moves which reduce the energy are allowed (falling). When the system is
blocked we tap it  with strength  $p\in [0,1/2]$, that is to say
each spin is flipped with a probability $p$,
the updating at this point being parallel. The system is
then evolved by the zero temperature dynamics until it becomes 
once again stuck, the tapping is then repeated. Physically this
corresponds to assuming that in granular media the relaxation time to
a new metastable state is much shorter than the time between taps.
The same tapping
dynamics has also been introduced independently 
in the context of three spin ferromagnetic
interactions on thin hypergraphs  \cite{mehta2}, also in the 
goal of studying the dynamics of granular media. We find that
a stationary regime  is reached
after a sufficiently large number of taps, characterized
by a steady state energy $E(p)$ (analogous to the stationary density
-- the same analogy as used in \cite{mehta2}).
We then develop a mean field theory
for the dynamics under falling then tapping, which 
appears to be exact in the case of the above mentioned one
dimensional system and one may calculate $E(p)$ within this scheme, the
results being in perfect agreement with the numerical simulations.
We also examine the tapping of spin glasses and ferromagnets
of higher connectivity. 
In the case of the ferromagnet we find numerically that there exists
a critical value $p_c$ of $p$ such that for $p > p_c$,
$E(p) > E_{GS}$ where $E_{GS}$ is the energy of the ground state and
the inequality is strict, and that for $p < p_c$ , $E(p) = E_{GS}$, hence
in the ferromagnetic system there is a first order phase transition
with the tapping dynamics (in contrast to the usual thermodynamic
ferromagnetic transition in these systems which is second order \cite{jon}). 
The spin glass 
is clearly far from a realistic realization of a granular media, however
the fact that it has  extensive entropy of blocked states and the 
obviously natural form of the tapping dynamics implemented makes it
a natural testing ground for ideas about dynamics and possible thermodynamics
of systems such as granular media. 
 
A random thin graph of connectivity $c$ 
is a collection of $N$ points, each point being linked to
$c$ of its neighbors.
The spin glass/ferromagnet model we shall consider has the Hamiltonian 
\begin{equation}
H = -{1\over 2} \sum_{j\neq i} J_{ij} n_{ij} S_i S_j
\end{equation} 
 where the $S_i$ are Ising spins, $n_{ij}$ is equal to one if the 
sites $i$ and $j$ are connected. In the spin glass case
the $J_{ij}$ are taken from a binary distribution where
$J_{ij}= -1$ with probability half and $  J_{ij}= 1$ with probability half.
In the ferromagnetic case $J_{ij} = 1$.
The definition of the total number of metastable states in this system is 
the number of states where any single spin flip does not increase the
energy of the system.

{\em The one dimensional spin glass/ferromagnet}:
We remark that by a gauge transformation
the  one dimensional ferromagnet and $\pm J$
spin glass are equivalent and place ourselves, 
for transparency, in the context of the ferromagnet. 
In the initial 
configuration,  we take the probability  
that a given spin is different to its left neighbor  to be $a$. Hence if 
$a=0$ we have an initially ferromagnetic configuration, if $a =1$ it
is an antiferromagnetic configuration, the case $a =1/2$ corresponds to
a completely random configuration. The initial energy per spin is  $E_0 = -1 + 2 a$.

For a given site
define $x$ to be the difference between the number of unsatisfied and 
satisfied bonds. Hence $x$ is the local field on the spin; 
if $x > 0$ then the spin can flip 
bringing about the change $x\to -x$. 
Denote by $P(x,k)$ the probability that
the site of interest has local field $x$ after a total of $k$ 
attempted random single spin flips during the falling dynamics.
We define $f_+$ and $f_-$ the probabilities that a neighboring spin 
can flip conditional
on the fact that the bond with the site we are interested in is not satisfied
or satisfied respectively. 
Given that a site has local field $x$, it must have $(c+x)/2$ unsatisfied bonds
and $(c-x)/2$ satisfied bonds; using Bayes' theorem we therefore obtain 
\begin{equation}
f_\pm  = {\sum_{ x> 0} P(x)(c\pm x) \over \sum_x P(x)(c\pm x)}
\end{equation}

For the spin considered the possibilities between 
time $k$ and $k+1$ are
\begin{itemize}
\item $x>0$. Then the spin can flip and $x$ goes to
$-x$;
\item  A neighboring spin has  a positive local field and
so can flip. In that case, $x$ goes to $x+2$ or to $x-2$;
\item Neither the spin considered nor any of
its neighbors flips, and so $x$ stays $x$. 
\end{itemize}
Assuming that the distribution at every site is given by $P(x,k)$ and 
assuming independence between the values of $x$ from site to site (the
mean field approximation) we obtain, taking the limit $N \to \infty$ and
 introducing the continuous time
$\tau = k/N$,
\begin{eqnarray}\label{eqn}
{dP(x)\over d\tau}&=& \theta(-x)P(-x)+\theta(-x)P(x)
-(c+1)P(x) \\ \nonumber
        &+&P(x)\,\left(\frac{c+x}{2}(1-f_+)
+\frac{c-x}{2}(1-f_-)\right)\\ \nonumber
	&+&P(x+2)\,\frac{c+x+2}{2}f_++P(x-2)\,\frac{c-x+2}{2}f_-
\end{eqnarray}
The case where $c=2$ is accessible to analytic 
solution. Defining: $u(\tau) \equiv P(-2,\tau)$,
$v(\tau) \equiv P(0,\tau)$ and $w(\tau) \equiv P(2,\tau)$, the solution
to Eq. (\ref{eqn}) is given by
\begin{eqnarray}
w(\tau)&=&w(0)\,\exp\left(-\tau+\frac{4}{\lambda(0)+2}(e^{-\tau}-1)\right)\\
v(\tau)&=&-2w(\tau) +w(0)(\lambda(0)+2)\,
\exp\left(\frac{4}{\lambda(0)+2}(e^{-\tau}-1)\right) \nonumber
\end{eqnarray}
where $\lambda(0) = v(0)/\omega(0)$. Consequently after the system has fallen 
into a metastable state we find the final values
$v(\infty)=(v(0)+2w(0))\,e^{-\frac{4}{\lambda(0)+2}}$, $\omega(\infty) = 0$
and $u(\infty) = 1- v(\infty)$.

For initial conditions used here: $u(0)=(1-a)^2$, 
$v(0)=2a(1-a)$ and $w(0)=a^2$. Hence $E_f$, the average energy of the 
metastable state into which the system falls, is given by

\begin{equation}
E_f=-1+v(\infty)=-1+2a e^{-2a}
\end{equation}
This result can be shown to
be exact by a combinatorial calculation and has also been checked
in our simulations \cite{pap2}.
For the completely random initial configuration,
where $a = 1/2$, $E_f= -0.632121$ is in fact maximal   

This calculation demonstrates two
important points:
\begin{itemize}
\item The final value of the energy $E_f$ depends strongly on the initial 
configuration, in addition $E_f$ is not a monotonic function of $E_0$. This
means that configurations of higher initial energy can fall into metastable 
states of lower energy than initial configurations of a lower energy.

\item The system does not fall into a state of energy corresponding
to the maximum of $N_{MS}(E)$, the total number of metastable states of 
energy $E$ per spin.
In \cite{dean2,lefde} it was shown that $N_{MS}(E) \sim \exp(N s_{Edw}(E))$
where $s_{Edw}(E)$ is a concave function
peaked at $E^* = -1/\sqrt{5} \approx 0.44721$. Hence even if the total number
of metastable states is dominated (in the thermodynamics sense)
by those of energy $E^*$, generic initial
conditions always seem to lead to an energy lower than this
\cite{jorge,mehta2}.  
\end{itemize}

Tapping the system with  probability $p$, starting from the 
values 
$\{u(\infty),v(\infty),w(\infty) \}$, we obtain the new tapped values 
$\{u^\prime(0),v^\prime(0),w^\prime(0) \}$. Defining $q \equiv
(1-p)$, the relations between the old and tapped probabilities are:

\begin{eqnarray}
u^\prime(0)&=&(1-3pq)\,u(\infty)+pq\,v(\infty) \\ \nonumber
v^\prime(0)&=&2pq\,u(\infty)+(1-2pq)\,v(\infty) \\ \nonumber
w^\prime(0)&=&pq
\end{eqnarray}
After another zero temperature evolution of the system, it reaches
a new local energy probability distribution with 
$w^\prime(\infty)=0$, $v'(\infty)$ and $u'(\infty) = 1-v'(\infty)$. The
fixed point of this dynamics is given by
\begin{equation}\label{eqn4}
v_{s}(p)=(4pq+v_{s}(p)(1-4pq))\,
\exp\left(-\frac{4pq}{4pq+v_{s}(p)(1-4pq)}\right)
\end{equation}
the subscript {\em s} indicating steady state.
We  remark that this solution 
contains one of the main features of our numerical simulations, that is
{\em reversibility}. Here the 
asymptotic distribution is independent of the initial conditions and depends
only on  $p$. 
Eq. (\ref{eqn4})
can be solved numerically and the result is shown in Fig. (1) in
comparison with the numerical simulations which we see is excellent.
The small $p$ behavior of $E(p)$ from (\ref{eqn4}) is:
$E(p)=-1+\sqrt{2p}+O(p)$.

{\em Systems with $c>2$:}
cases. 
In 
\cite{dean2,lefde} an annealed approximation to the Edwards entropy per spin of
metastable states at fixed energy $E$, 
$s_{Edw}(E) = \ln\left( \langle N_{MS}(E)\rangle\right)/N$  was carried out
for spin glasses and ferromagnets on random thin graphs. 
There is an energy threshold $E^*$ above which the results
are the same for the $\pm J$ spin glass and the ferromagnet;
below $E^*$  the ferromagnet has more metastable
states (which aquire a non-zero magnetization \cite{lefde}). 
Hence, as far as metastable states are
concerned, both ferromagnet and spin glass are the same above $E^*$, that is
the effect of loop frustration is negligible. In this regime, one  suspects
that the zero temperature dynamics are the same. In
particular, numerical simulations with $100$ samples of $N=10000$ sites
for connectivities of $3$, $4$ and $5$
have found the same $E_f$ for the spin glass and ferromagnet
to very good accuracy (the relative error is
about $10^{-6}$). The results of tapping experiments on 
systems with $c=3$ are 
displayed in Fig.(2). There is a critical tapping rate $p_c$ above
which the curves of $E(p)$ versus $p$ are the same
for the spin glass and ferromagnet. At $p_c \approx 0.249$ the ferromagnet
undergoes a phase transition  such
that for $p<p_c$, the steady state reached is the ground state.  
Finite size effects have been studied and reveal that the transition 
appears to  be first order (in as far that $E(p_c^+) \neq E(p_c^-)$).
As shown  in  Fig.(3), there is a coexistence of two
phases at $p_c$; one sees  two separated
peaks in the distribution of the internal energy 
and not a single peak  as one 
would expect for a second order transition. 
In \cite{lefde} it was
shown that for the ferromagnet the annealed approximation to the
Edwards entropy as a function of $E$
is concave for $E>E^*$ and convex for $E<E^*$. The value of $E(p_c^+)$
estimated  from the tapping experiments are very close to those 
obtained for $E^*$ in \cite{lefde}, the energy at which $s_{Edw}(E)$ becomes 
convex (for $c=3$, $E^* = -1.0714$ analytically
and $E(p_c^+) \approx -1.075\pm 0.005$
from the simulation).
Encouraged by this
observation we will try to make a tentative link with a 
possible thermodynamics for such systems. We consider 
a partition function inspired by the 
flat Edwards measure over metastable states \cite{edw,nico} $
Z = \int dE N_{MS}(E) \exp(-\beta N E)$
where $\beta$ is a Lagrange multiplier corresponding to the inverse Edwards
temperature which depends solely on $p$, and not on $E$, and is a monotonically
decreasing function of $p$ for $p \in [0^+, 1/2]$. The monotonicity 
hypothesis is supported by the simulation results that $E(p)$ decreases with
decreasing $p$. Clearly the energy which dominates in the sum is that obeying
${\partial s_{Edw}(E)\over \partial E} -\beta = 0$. 
However if this saddle point gives a true maximum of the action 
the Edwards entropy must be concave for the energy
considered to be thermodynamically stable. Hence one is lead to conclude that
for $E< E^*$ the only stable energy is the ground state. 
Finally we remark  for $c \geq 3$ that the 
mean field equations of the previous section
may be numerically solved and give reasonable results for $E(p)$ for 
$p > p_c$ \cite{pap2}, 
which again reinforces the assertion that for higher energies the
spin glass and ferromagnetic dynamics are the same
(the mean field theory does not distinguish between the two)..

No observable singularities are seen in our simulations of spin glasses,
but again we found that for all the systems studied that $E(p)$ decreased
as a function of $p$. As $p \to 0$ we found that generically
$E(p) \sim E(0^+) + A p^\theta$. For the one dimensional case
we know analytically that $\theta = 1/2$, however for $c= 3,\ 4,$ and $5$ 
we found that  $\theta=1$. We also
carried out simulations on the totally connected Sherrington Kirkpatrick 
$\pm J$ spin glass for systems of size $400,\ 600$ and $800$ spins 
where an average was taken over $1000$ samples. Here we used a 
deterministic sequential (rather than random) update for the zero temperature
dynamical evolution -- as used by Parisi in \cite{par} where the numerical
evaluation of $E_f$ was undertaken. Interestingly we found that the finite 
size scaling introduced by Parisi for $E_f$ worked extremely well for 
$E(p)$ that is $E(p, \infty) \sim E(N,p) - 1/4N^{1\over 3}$, for all $p$. 
In addition
we found that for $p$ small $E(p)- E(0^+) \sim A p^\theta$ with 
$\theta \approx 0.4 \pm 0.1$ \cite{pap2}. We emphasize  that all of
the numerical simulations found here (for spin glasses and ferromagnets) 
exhibit perfect
reversibility for $E(p)$. In addition the values of $E_f$ found in all
systems were less than the value of the energy dominating thermodynamically
in $s_{Edw}(E)$ (calculated in \cite{dean2,lefde}), 
showing that the falling dynamics in strongly non ergodic
\cite{mehta2}.
 
Granular media are examples of systems having an extensive entropy
of metastable states. In such systems the role of thermal fluctuations
are negligible and in order to evolve one must apply some external
tapping mechanism. One would ultimately like to be able to formulate
some sort of thermodynamics for such systems. The proposition of 
Edwards \cite{edw}
for such a thermodynamics is an important step in this
direction and has had some success \cite{jorge} but in the same paper
it was shown not generically to be true. A more
general understanding of the asymptotic states of tapped systems has
also far reaching implications for computer science as the tapping
mechanism studied here is similar to certain algorithms used in
optimization problems. We have presented what appears to be the 
exact solution to the problem of tapping a one dimensional 
spin glass or ferromagnet. 
The fixed point equation for Eq.(\ref{eqn4})
may have a thermodynamic interpretation
which presents an open challenge to find. In a wide context of models
we confirm the observations of \cite{exps,mehta1,mehta2}, that if one reduces
the {\em strength} of tapping, then the compaction process, corresponding
here to the reduction of the energy of the system, becomes more efficient.
The existence of a first order type phase transition for tapped
ferromagnets on random thin graphs is of great interest, the 
possible explanation using the calculations of \cite{lefde} on the 
Edwards entropy for this system indicates the possibility of
constructing a  general theory for the thermodynamics
and  phase transitions in tapped systems.

\end{multicols}

{\bf {Figure Captions}}

Fig. 1.  Comparison between numerical simulations of tapping 
         experiments (b) and the analytical result (a) obtained
         with (\ref{eqn4}).

Fig. 2.  Numerical simulations of tapping experiments for the spin glass (c)
         and the ferromagnet ((a), (b) and (d)) for $c=3$ for $N=1000$ 
         ((c) and (d)), $N=2000$ (b) and $N=10000$ (a). The inset shows 
         the scaling $E^{(N)}=f(N(p-p_c))$ for $p \simeq p_c$ for $N=400$, 
         $N=1000$ and $N=2000$.

Fig. 3.  Histogram of the distribution $p(E)$ of energy during a tapping
	 simulation at $p \approx p_c$ for a single run of $50000$ taps and
	 $N=5000$ spins. 
  
\begin{multicols}{2} 
  
\begin{figure}
\narrowtext
\epsfxsize=0.8\hsize
\epsfbox{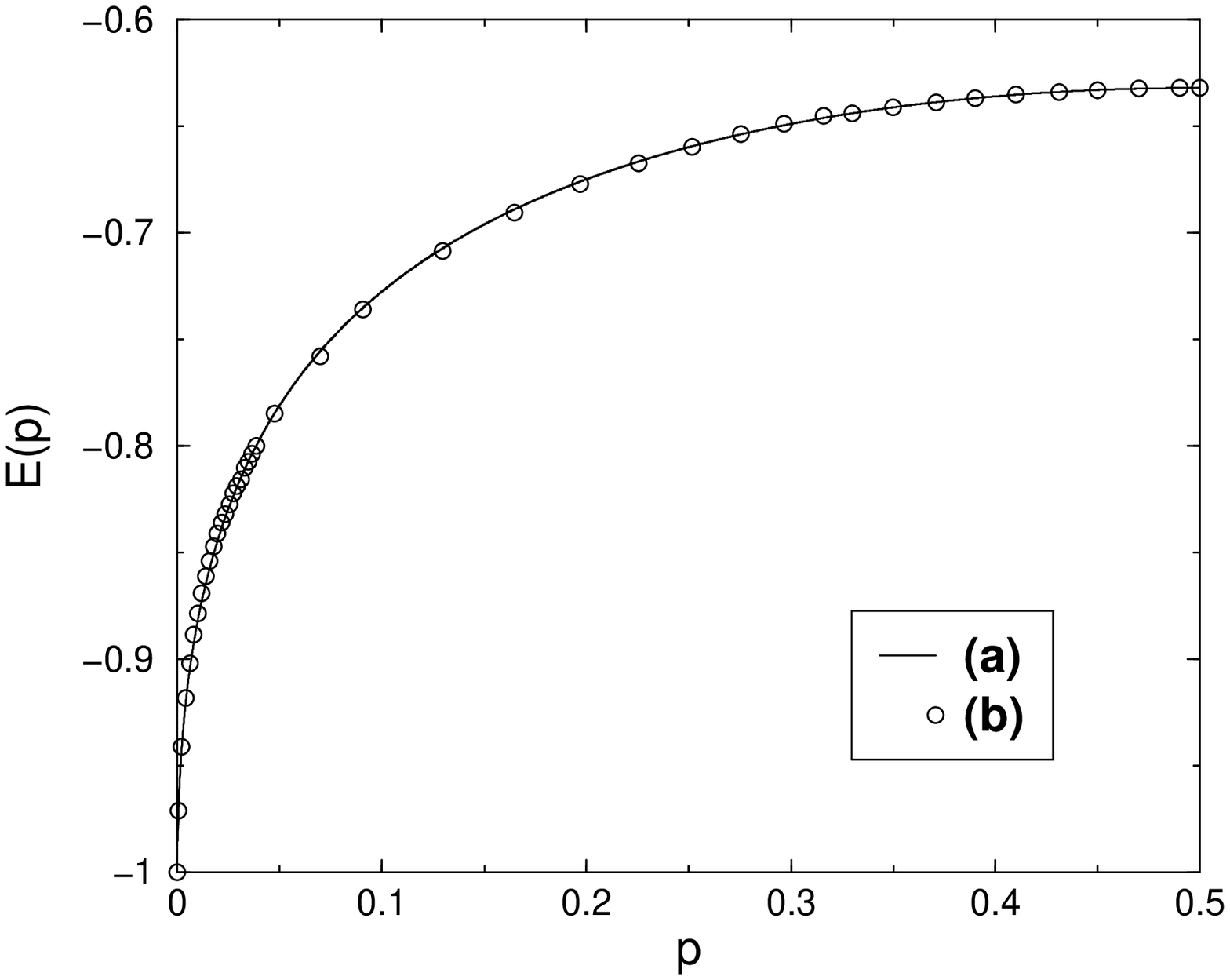}
\label{fig1}
\end{figure}

\begin{figure}
\narrowtext
\epsfxsize=0.8\hsize
\epsfbox{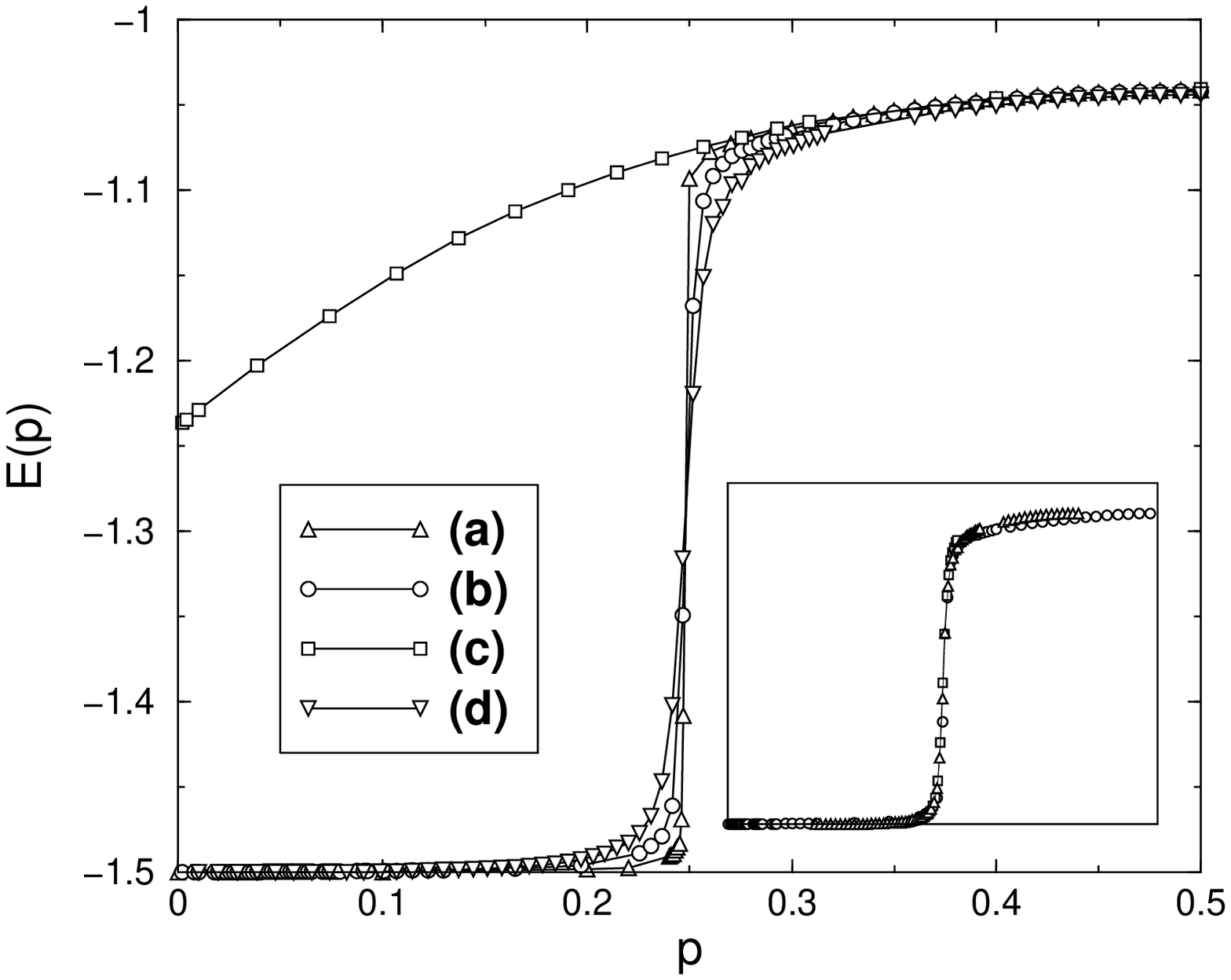}
\label{fig2}
\end{figure}

\begin{figure}
\narrowtext
\epsfxsize=0.8\hsize
\epsfbox{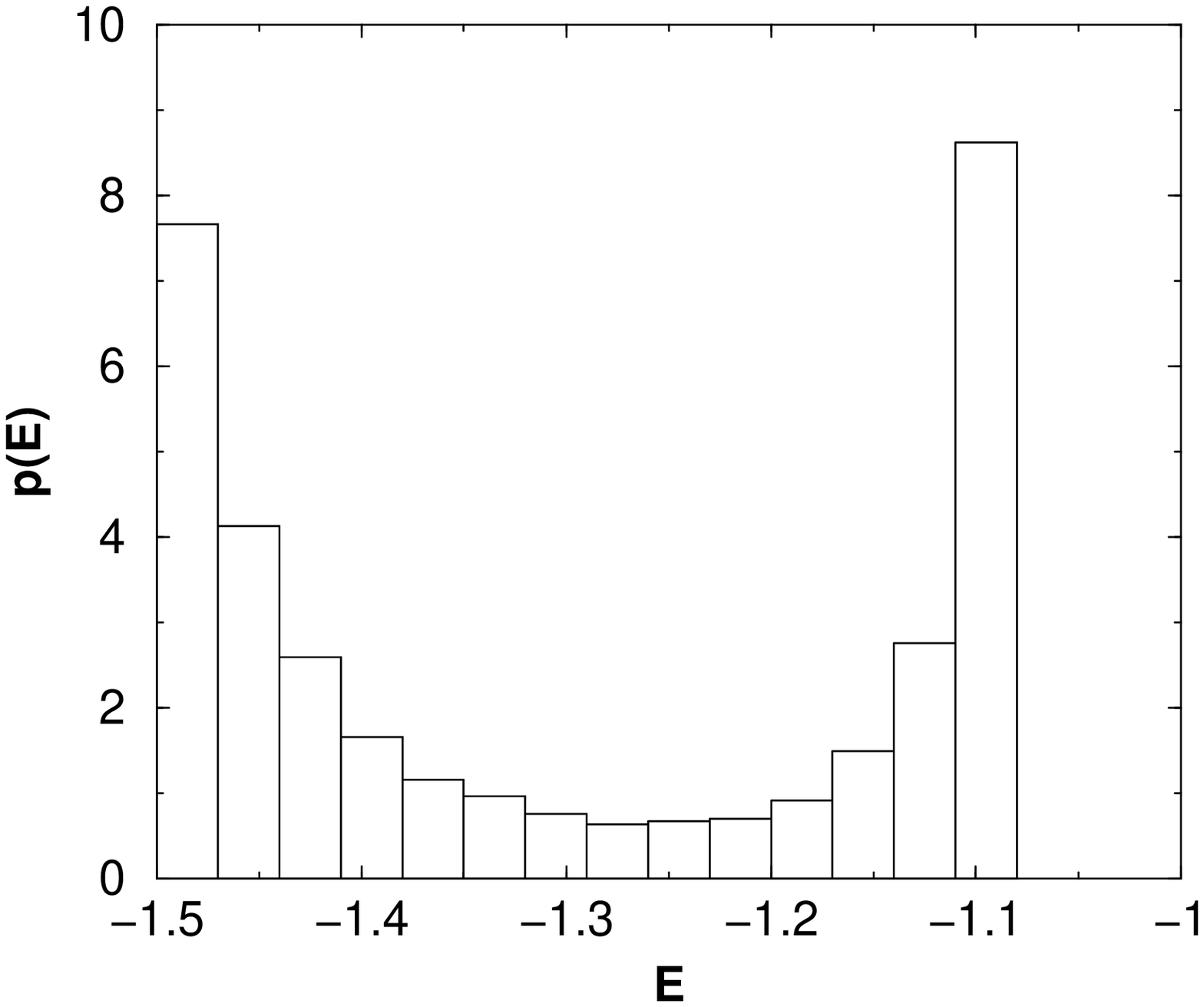}
\label{fig3}
\end{figure}
\end{multicols}

\end{document}